\documentclass[a4paper]{aipproc}
\layoutstyle{8x11single}

\newcommand{\flr}[3]{\mathopen{}\lr#1{#2}#3\mathclose{}}
\newcommand{\lr}[3]{\left#1#2\right#3}
\renewcommand{\d}[0]{{\rm d}}

\begin{document}

\title[Photon emission by ultra-relativistic positrons
in crystalline undulators]
{Photon emission by ultra-relativistic positrons
in crystalline undulators: the high-energy regime.
\footnote{
submitted for the proceedings of the International
Workshop on ``Electron-Photon Interaction in Dense Media'' in
Nor-Hamberd, Armenia, 2001
}
}

\author{Wolfram Krause\footnote{E-mail: {\tt krause@th.physik.uni-frankfurt.de}} }{
	address={Institut f\"ur Theoretische Physik der Johann Wolfgang
	Goethe-Universit\"at, 60054 Frankfurt am Main, Germany},
	email={krause@th.physik.uni-frankfurt.de},
}

\author{Andrei V.\ Korol\footnote{E-mail: {\tt korol@th.physik.uni-frankfurt.de}\\Department of Physics, St.Petersburg
State Maritime Technical University, Leninskii prospect 101,
St. Petersburg 198262, Russia} }{
	address={Institut f\"ur Theoretische Physik der Johann Wolfgang
	Goethe-Universit\"at, 60054 Frankfurt am Main, Germany},
,
	email={korol@th.physik.uni-frankfurt.de},
}

\author{Andrey V.\ Solov'yov\footnote{E-mail: {\tt solovyov@th.physik.uni-frankfurt.de}\\
A.F.Ioffe Physical-Technical Institute, Russian Academy of Sciences,
Polytechnicheskaya 26, St. Petersburg 194021, Russia} }{
	address={Institut f\"ur Theoretische Physik der Johann Wolfgang
	Goethe-Universit\"at, 60054 Frankfurt am Main, Germany},
	email={solovyov@th.physik.uni-frankfurt.de},
}

\author{Walter Greiner\footnote{E-mail: {\tt greiner@th.physik.uni-frankfurt.de}} }{
	address={Institut f\"ur Theoretische Physik der Johann Wolfgang
	Goethe-Universit\"at, 60054 Frankfurt am Main, Germany},
	email={greiner@th.physik.uni-frankfurt.de},
}







\date{\today}

\begin{abstract}
This paper discusses the undulator radiation emitted by high-energy
positrons during planar channeling in periodically bent crystals.  We
demonstrate that the construction of the undulator for positrons with
energies of 10 GeV and above is only possible if one takes into
account the radiative energy losses.  The frequency of the undulator
radiation depends on the energy of the particle. Thus the decrease of
the particle's energy during the passage of the crystal should result
in the destruction of the undulator radiation regime.  However, we
demonstrate that it is possible to avoid the destructive influence of
the radiative losses on the frequency of the undulator radiation by
the appropriate variation of the shape of the crystal channels. We
also discuss a method by which, to our mind, it would be possible to
prepare the crystal with the desired properties of its channels.
\end{abstract}

\pacs{41.60}

\maketitle


\section{Introduction}
\label{sec:intro}

We discuss a mechanism, initially proposed in
\cite{Korol98,Korol99}, for the generation of high-energy photons
by means of planar channeling of ultra-relativistic positrons
through a periodically bent crystal. In this system there appears, in
addition to the well-known channeling radiation, an undulator type
radiation due to the periodic motion of the channeling positrons which
follow the bending of the crystallographic planes. The intensity and
the characteristic frequencies of this undulator radiation can be
easily varied by changing the positrons energy and the parameters of
the crystal bending.

\begin{figure}
\includegraphics{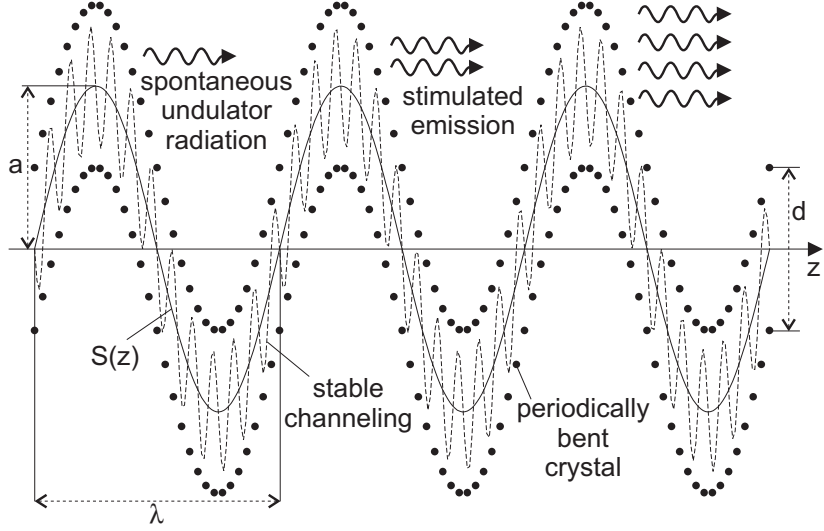}
\caption{Schematic figure of the crystalline undulator. The scale in
$y$ direction is magnified by a factor $> 10^4$. The function $S(z)$
defines the shape of the centerline of the periodically bent channel.}
\label{fig:undulator}
\end{figure}

The mechanism of the photon emission by means of the crystalline
undulator is illustrated in figure \ref{fig:undulator}.  It is
important to stress that we consider the case when the amplitude
$a$ of the bending is much larger than the interplanar spacing $d$
($\sim 10^{-8}\;\mathrm{cm}$) of the crystal ($a \sim 10\ d$), and,
simultaneously, is much less than the period $\lambda$ of the
bending ($a \sim 10^{-5} \dots 10^{-4}\, \lambda$).

In addition to the spontaneous photon emission the scheme leads to the
possibility to generate stimulated emission. This is due to the fact,
that the photons emitted at the points of maximum curvature of the
trajectory travel almost parallel to the beam and thus, stimulate the
photon generation in the vicinity of all successive maxima and minima
of the trajectory.

The bending of the crystal can be achieved either dynamically or
statically. In \cite{Korol98,Korol99} it was proposed to use a
transverse acoustic wave to bend the crystal dynamically. The
important feature of this scheme is that the time period of the
acoustic wave is much larger than the time of flight of a bunch of
positrons through the crystal and thus the crystal bending does not
change on this time scale. One possibility to create acoustic
waves in a crystal is to place a piezo sample atop the crystal and
to use radio frequency to excite oscillations.

The usage of a statically and periodically bent crystal was discussed
in \cite{Uggerhoj00}. The idea is to construct a crystalline undulator
based on graded strained layers. We will present a detailed
description how a static crystalline undulator can be produced.

We now consider the conditions for stable channeling. The channeling
process in a periodically bent crystal takes place if the maximum
centrifugal force in the channel, $F_{\mathrm{cf}}\approx m \gamma
c^2/R_{\mathrm{min}}$ ($R_{\mathrm{min}}$ being the minimum curvature
radius of the bent channel), is less than the maximal force due to the
interplanar field, $F_{\mathrm{int}}$ which is equal to the maximum
gradient of the interplanar field (see \cite{Korol99}). More
specifically, the ratio $C=F_{\mathrm{cf}}/F_{\mathrm{int}}$ has to be
smaller than 0.15, otherwise the phase volume of channeling
trajectories is too small (see also \cite{Korol00}). Thus, the
inequality $C<0.15$ connects the energy of the particle, $\varepsilon=m
\gamma c^2$, the parameters of the bending (these enter through the
quantity $R_{\mathrm{min}}$), and the characteristics of the
crystallographic plane.

A particle channeling in a crystal (straight or bent) undergoes
scattering by electrons and nuclei of the crystal. These random
collisions lead to a gradual increase of the particle energy
associated with the transverse oscillations in the channel. As a
result, the transverse energy at some distance $L_d$ from the
entrance point exceeds the depth of the interplanar potential well,
and the particle leaves the channel. The quantity $L_d$ is called the
dechanneling length \cite{Gemmel74}. To calculate $L_d$ one may follow
the method described in \cite{Biryukov96,Krause00a}. Thus, to
consider the undulator radiation formed in a crystalline undulator, it
is meaningful to assume that the crystal length does not exceed $L_d$.

In \cite{Krause00a} we estimated the parameters $a$ and $\lambda$ for
given energy $\varepsilon$, regarding the dechanneling length of the
bent crystal and the reduction of the phase-space volume due to the
bending.  For 500 MeV positrons in Si(110) the optimal parameters are
$a/d=10$ and $\lambda = 2.335 \cdot 10^{-3}$ cm. The spectral
distribution of the emitted radiation in this case is discussed in
the next section (see also \cite{Krause00a}).

In the present paper we discuss the possibility to construct
undulators to generate photons with energies larger than 1 MeV using
positron energies above 10 GeV when the radiative energy losses cannot
be neglected and, thus, must be taken into account \cite{Korol00}.

The frequency of photons generated in the undulator is determined by
the energy of the projectiles and also by the undulator parameter (for
definition see equation (\ref{losses_comp_1})). In the regime in which
the energy of the projectiles is not constant during their passage
through the undulator, the frequency of the emitted undulator
radiation can nevertheless be kept constant if one chooses the
appropriate variation of the shape of the undulator along its length.

We also discuss a method by which, to our mind, it would be possible
to prepare crystals with the desired properties of their channels.

\section{Spectra of the spontaneous emitted radiation}
\label{sec:spectra}

To illustrate the undulator radiation phenomenon, which we discuss,
let us consider the spectra of spontaneous radiation emitted during
the passage of positrons through periodically bent crystals.

The photon emission spectra have been calculated using the
quasiclassical method \cite{Baier98}.  The trajectories of the
particles were calculated numerically and then the spectra were
evaluated \cite{Krause00a}. The latter include both radiation
mechanisms, the undulator and the channeling radiation.

\begin{figure}
\includegraphics{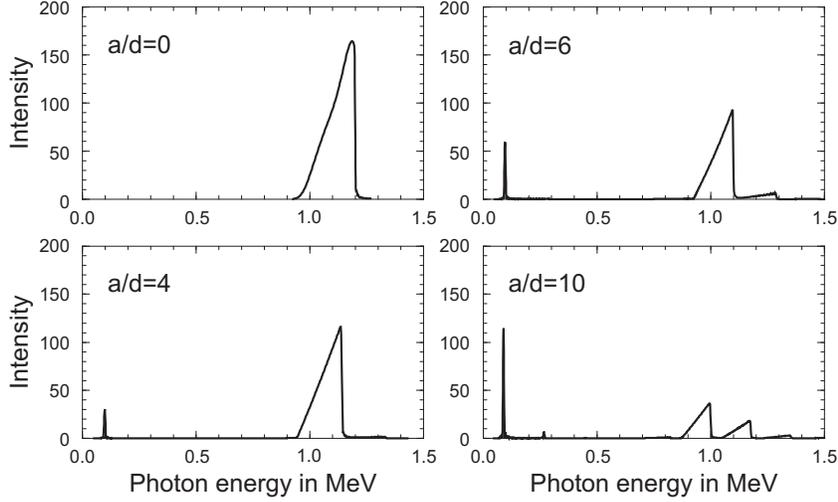}
\caption{Spectral distributions of the total radiation emitted in
forward direction for $\varepsilon=500$ MeV positrons channeling in Si
along the (110) crystallographic planes for different $a/d$
ratios.}
\label{fig:spectra_500mev}
\end{figure}

The spectral distributions of the total radiation emitted in forward
direction for $\varepsilon=500$ MeV positrons channeling in Si along
the (110) crystallographic planes are plotted in figure
\ref{fig:spectra_500mev}. The wavelength of the crystal is fixed at
$\lambda=2.335\cdot10^{-3}$ cm, while the ratio $a/d$ is changed from
0 to 10. The length of the crystal is $L_d=3.5\cdot10^{-2}$ cm and
corresponds to $N=15$ undulator periods.

The first graph in figure \ref{fig:spectra_500mev} corresponds to the
case of the straight channel ($a/d=0$) and, hence, presents the
spectral dependence of the ordinary channeling radiation
only. Increasing the $a/d$ ratio leads to modifications in the
spectrum of radiation.  The changes which occur manifest themselves
via three main features: the lowering of the ordinary channeling
radiation peak, the gradual increase of the intensity of
undulator radiation due to the crystal bending and the appearing of
additional structure (the sub-peaks) in the vicinity of the first
harmonic of the ordinary channeling radiation.  A more detailed
analysis of these spectra can be found in \cite{Krause00a}.

\begin{figure}
\includegraphics{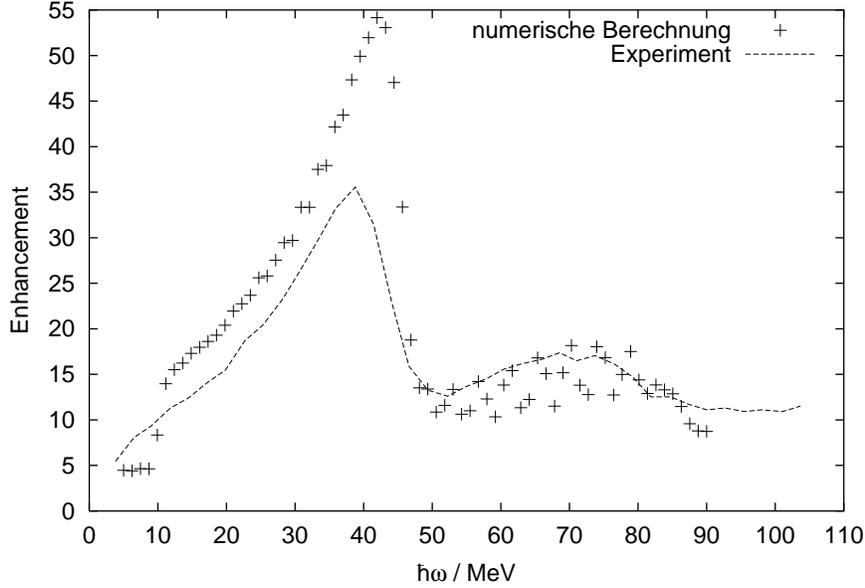}
\caption{Comparison of the experimentally  measured spectrum
\cite{Bak85,Uggerhoj93} and the results of our calculation for 6.7 GeV
positrons in Si(110).}
\label{fig:spectrum67}
\end{figure}

To check our numerical method, we have
calculated the spectrum of the pure channeling radiation for 6.7 GeV
positrons in Si(110) integrated over the emission angles. Figure
\ref{fig:spectrum67} shows the experimental data
\cite{Bak85,Uggerhoj93} and the results of our calculations,
normalized to the experimental data in the vicinity of the second harmonic.

The energy and the spectral dependence of the calculated spectra is in
good agreement with the experimental data. The fluctuations at high
energies are an artifact of our numerical method. Increasing the
number of calculated trajectories will reduce these fluctuations but
also increase the computation time.

The height of the first harmonic is overestimated in our
calculations. The calculations performed in \cite{Bak85} give a
similar result. This disagreement arises likely due to the neglection
of multiple collisions both in our work and in \cite{Bak85}. The shape
and the location of the first harmonic are described quite well. This
fact demonstrates that the Moli\`ere potential is a good approximation for
the interplanar potential, because the spectral distribution of the
channeling radiation is highly sensitive to the shape of the
interplanar potential.

\section{Undulator effect in the high-energy regime}
\label{cha:losses}

Spectra of channeling and undulator radiation presented in the
previous section have been calculated in the regime in which the energy
losses of the positrons during their passage through the crystal are
negligible. In this section, we analyze the opposite situation, which
occurs when the energy of the projectiles becomes sufficiently large
(above 10 GeV). On the first glance, the undulator phenomenon can
hardly take place in this energy range, because the energy of positrons
during their passage through the crystal can no longer be considered
as constant due to the radiative energy losses \cite{Korol00}.

Indeed, the frequency $\omega_{\mathrm{und}}^{(1)}$ of the first
harmonic of the undulator radiation in the forward direction is given
by \cite{Korol99,Korol01a}:
\begin{equation}
\label{losses_comp_1}
\omega_{\mathrm{und}}^{(1)}=
\frac{4\,\omega_0\,\gamma^2}{2+p_{\mathrm{und}}^2}
= \frac{4 \pi\, c \, \gamma^2(z)}{\lambda+2\pi^2 \,\frac{a^2}{\lambda}\,
\gamma^2(z)}.
\end{equation}
Here we use $\omega_0=2\pi\,c/\lambda$ and the undulator parameter
$p_{\mathrm{und}}$ is defined as $p_{\mathrm{und}}= \gamma \, 2 \pi\,
a/\lambda$. The shape of the crystal is $S(z)=a\, \sin(kz)$.

Equation (\ref{losses_comp_1}) shows that the frequency of the emitted
radiation depends on the energy of the projectile. If the decrease of
the particle's energy due to the radiative losses is significant
($\gamma(z) < \gamma_0$ for $z>0$), the frequency
$\omega_{\mathrm{und}}^{(1)}$ becomes dependent on the particle's
penetration distance $z$ into the crystal. The decrease of the
particle's energy leads to the broadening of the undulator lines in
the photon emission spectrum and the reduction of their intensity.

However, the monochromaticity of the undulator radiation in the
high-energy regime can be restored if one allows the variation of the
shape of the crystal channels. Let us consider this condition in more
detail and assume that the shape of the channels in the crystal is as
follows:
\begin{equation}
\label{losses_comp_3}
S(z)=a(z) \, \sin\flr({\varphi(z)})
\end{equation}
with $\varphi(z)=\int_0^z\, 2\pi/\lambda \, \d z$,  $a(z)$ and
$\lambda(z)$ are the amplitude and the ``period'' of the bent
crystal channels as function of the penetration depth
$z$.

Let us formulate the conditions for the choice of the shape function
$S(z)$. For the given dependence $\gamma(z)$ the functions $a(z)$ and
$\lambda(z)$ have to be chosen to keep constant the frequency of the
first harmonics, $\omega_{\mathrm{und}}^{(1)}(z) =
\mathrm{const.}$ In addition, we require $C(z)=C=\mathrm{const.}$ It
was shown in
\cite{Krause00a,Korol01a}, that the parameter $C$ is the essential
characteristic for the channeling process in bent channels and the
regime in which the process happens. The parameter $C$ is defined
through \cite{Korol99}:
\begin{equation}
\label{losses_comp_5}
C =
\frac{\varepsilon(z)}{R_{\mathrm{min}}(z)\,U_{\mathrm{max}}^\prime}
=
\frac{4 \pi^2\, m c^2}{U_{\mathrm{max}}^\prime}\,
\frac{a(z)}{\lambda^2(z)}\, \gamma(z).
\end{equation}
Here $R_{\mathrm{min}}\approx \lambda^2(z)/(4\pi\,a(z))$ is the
curvature radius of the shape function $S(z)$ in the points of its
extrema. The formula, connecting $R_{\mathrm{min}}$, $\lambda(z)$ and
$a(z)$, is written as for the pure sine function with constant $a(z)$
and $\lambda(z)$. This can be done, because the parameters $a(z)$ and
$\lambda(z)$ change slowly with increasing $z$ and can be assumed to
be constant on the length of a single undulator period. This
assumption allows one to describe the bent channel locally by the sine
function. Rewriting equation (\ref{losses_comp_5}), we derive the
following expression for the amplitude $a(z)$:
\begin{equation}
\label{losses_comp_6}
a(z) =
\frac{\lambda^2(z)}{\gamma(z)}\,
\frac{C\, U^\prime_{\mathrm{max}}}{4 \pi^2\, mc^2}.
\end{equation}
Substituting (\ref{losses_comp_6}) in (\ref{losses_comp_1}), one
derives the following cubic equation for $\lambda(z)$:
\begin{equation}
\label{losses_comp_8}
\lambda^3(z) + a_1 \, \lambda(z) + a_0 = 0
\end{equation}
with the coefficients
\begin{equation}
\label{losses_comp_9:0}
a_0 = -32\pi^3\,c\,
\frac{1}{\omega_{\mathrm{und}}^{(1)}}
\,
\lr({\frac{\gamma(z)\,mc^2}{C\,U^\prime_{\mathrm{max}}}})^2
\quad \mbox{and} \quad
a_1 = 
\frac{8 \pi^2\,\lr({mc^2})^2}{C^2\,U^\prime_{\mathrm{max}}}.
\end{equation}
According to \cite{Abramowitz}, the real solution of equation
(\ref{losses_comp_8}) reads:
\begin{equation}
\lambda(z) =
\lr({- \frac{a_0}{2}+\sqrt{\frac{a_1^3}{27}+\frac{a_0^2}{4}}})^{1/3}
+ \lr({- \frac{a_0}{2}-\sqrt{\frac{a_1^3}{27}+\frac{a_0^2}{4}}})^{1/3}.
\label{losses_comp_11}
\end{equation}

Equations (\ref{losses_comp_6}) and (\ref{losses_comp_11}) contain the
dependence $\gamma(z)$ which describes the decrease of the particle's
energy in the crystal due to the radiative energy losses. For
comparatively low energies of the projectile ($\varepsilon < 10 GeV$)
this dependence can be calculated using the approach suggested in
\cite{Korol00}. To describe the radiative losses of particles in the
high-energy regime, one has to modify the formulas outlined in
\cite{Korol00}. Namely, it is necessary to replace the dechanneling length
$L_{\mathrm{d}}$ by the infinitesimal interval $\d z$ and also use
infinitesimal intervals for the energy loss. Physically, this means
that in the high-energy regime the particle's energy changes over
distances which are much smaller than the dechanneling length.  Thus,
the dependence of $\gamma$ on the penetration depth into the crystal
is given by:
\begin{equation}
\label{losses_sc_1}
\frac{\d \gamma}{\d z} = -
2.3\cdot10^{-23} \, \frac{(U_0/\mathrm{eV})^2}{(d/\mathrm{cm})^2} \;
G\flr({\gamma,C}) \; \gamma^2 \; \frac{1}{\mathrm{cm}}.
\end{equation}

The definition of $G\flr({\gamma,C})$ and the related details can be
found in \cite{Korol00}. $G$ includes the averaging over all possible
trajectories of the channeled particles.

\begin{figure}
\includegraphics{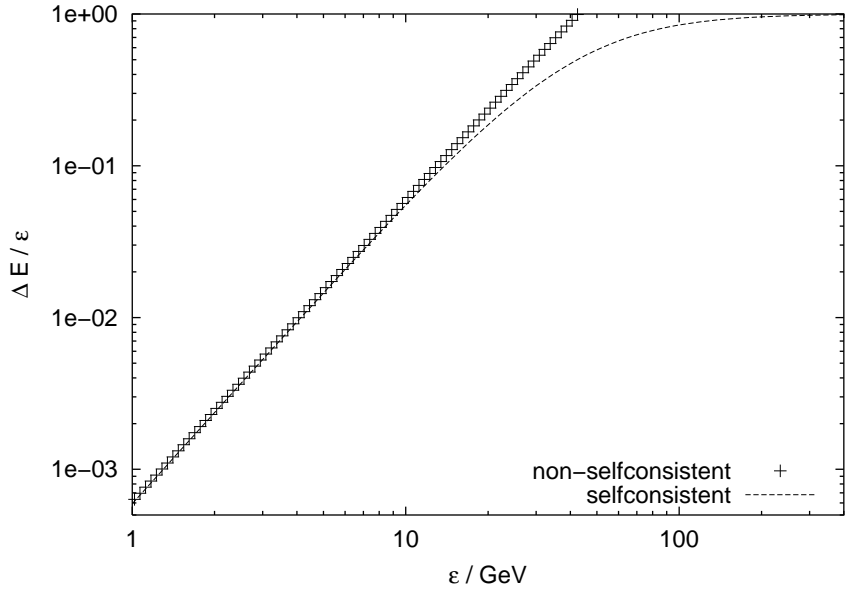}
\caption{Comparison of the fixed-energy and self-consistent
calculations of the radiative energy losses for positrons in Si(110),
$C=0.15$. See also explanations in the text.}
\label{fig:losses_selfconsistent}
\end{figure}

Solving (\ref{losses_sc_1}) numerically over the $z$-interval equal to
the dechanneling length one obtains the radiative losses. The result
of this calculation for $C=0.15$ is shown in figure
\ref{fig:losses_selfconsistent}. For the sake of comparison, we also
plot the dependence of the radiative energy losses in the low energy
regime \cite{Korol00}. As expected, the self-consistent losses grow up
slower at large energies and for $\varepsilon > 100$ GeV the losses
saturate at 1. For energies below 15 GeV the difference between the
two approaches is negligible small.  The absolute values of the
radiative loss become negligible for positron energies below 5 GeV,
which corresponds to the results derived in \cite{Korol00}.

Thus, starting from (\ref{losses_sc_1}) and the initial values
$\varepsilon_0$, $a_0$ and $\lambda_0$, one can calculate the energy
$\varepsilon(z)$ as a function of the penetration distance $z$.
Equations (\ref{losses_comp_6}) and (\ref{losses_comp_11}) allow then
the derivation of $\lambda(z)$ and $a(z)$. The ansatz
(\ref{losses_comp_3}) then determines the shape $S(z)$ of the channel.
The latter, in turn, ensures that the
frequency of the undulator radiation and the parameter $C$ remain
constant during the passage of the positrons through the crystal, even
in the regime in which the radiative energy losses are high. We
consider the possibility of the construction of such bent crystals in
the next section.

\begin{figure}
\includegraphics{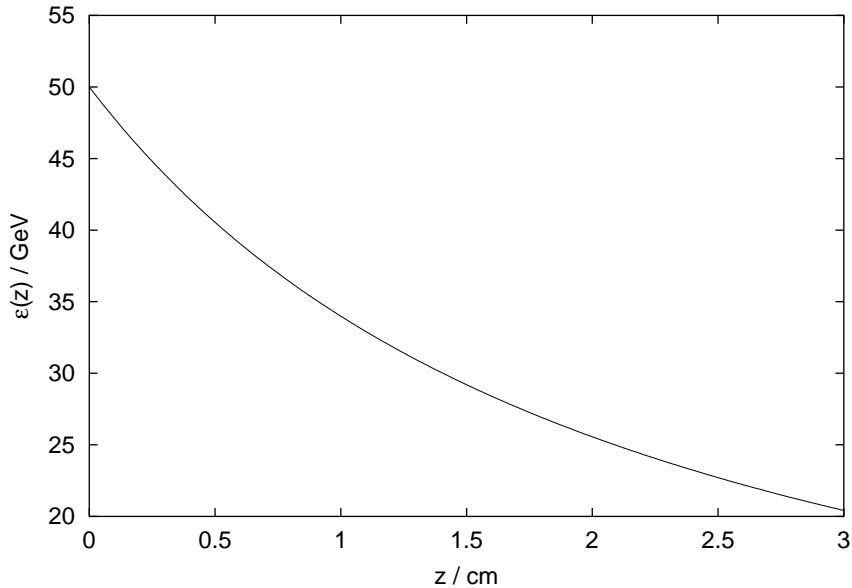}
\caption{The energy of the positrons as function of the penetration
depth $z$ in the high-energy channeling regime for Si(110) and initial
positron energy 50 GeV. The averaging over the possible initial
conditions of the positrons was performed as described in \cite{Korol00}.}
\label{fig:losses_epsilon}
\end{figure}

\begin{figure}
\includegraphics{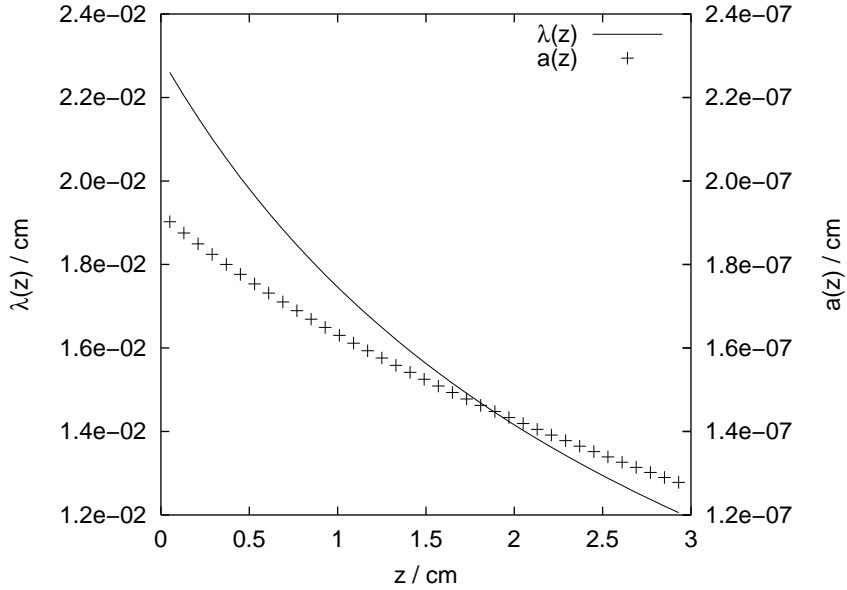}
\caption{The wavelength $\lambda$ and the amplitude $a$ of the bent
crystal as function of the penetration depth $z$ calculated according
to (\ref{losses_comp_11}) and (\ref{losses_comp_6}).}
\label{fig:losses_lambda}
\end{figure}

To illustrate the described method we consider positrons with
an initial energy of 50 GeV channeling in Si(110). Figures
\ref{fig:losses_epsilon} and \ref{fig:losses_lambda} show the results
of the calculations.

Figure \ref{fig:losses_epsilon} presents the energy of positrons
as a function of the penetration depth calculated by solving equation
(\ref{losses_sc_1}). We have chosen $C=0.15$ and the initial amplitude
$a_0= 10\, d$. These relationships define $\lambda_0 =
\sqrt{\varepsilon \, \pi^2\, a\, d/(U_{\mathrm{max}}^\prime\, C)} =
2.25\cdot 10^{-2}$ cm. The argumentation for the choice of $C$ and $a$
one finds in \cite{Korol00} and \cite{Krause00a}.

Using (\ref{losses_comp_6}) and (\ref{losses_comp_11}) we have
calculated the parameters $\lambda(z)$ and $a(z)$ characterizing the
shape of the channels.  The results are presented in figure
\ref{fig:losses_lambda}.  Having derived $a(z)$ and $\lambda(z)$, one
can easily calculate the shape of the channels using equation
(\ref{losses_comp_3}).

The particle density of channeling positron beams decreases exponentially
along the channel \cite{Biryukov96,Korol01a}.  The dechanneling length
for positrons of $\varepsilon=50$ GeV and $C=0.15$ is approximately
1.5 cm and the number of undulator periods on this length is about
75. The emitted undulator radiation should have
high intensity and narrow spectral width. The energy of photons at
the first harmonic emitted in the forward direction is
$\hbar\omega_{\mathrm{und}}^{(1)}=7.1$ MeV and the spectral width can
be estimated as $\Gamma_{\mathrm{und}}^{(1)}/2 =
\hbar\omega_{\mathrm{und}}^{(1)} / N_{\mathrm{und}} = 44$ keV.

\section{Growing of crystals with periodically bent channels}
\label{cha:crystal}

In this section we propose a method of preparing crystals with
periodically bent channels whose shape function $S(z)$ has either the
pure sine form, $a\, \sin kz$, or a more general one defined by
(\ref{losses_comp_3}).

In \cite{Breese97c} the deflection of proton beams by means of
strained crystal layers was demonstrated. The construction of the
crystals was described and experimental data that proves the deflection
of protons was presented.

\begin{figure}
\includegraphics{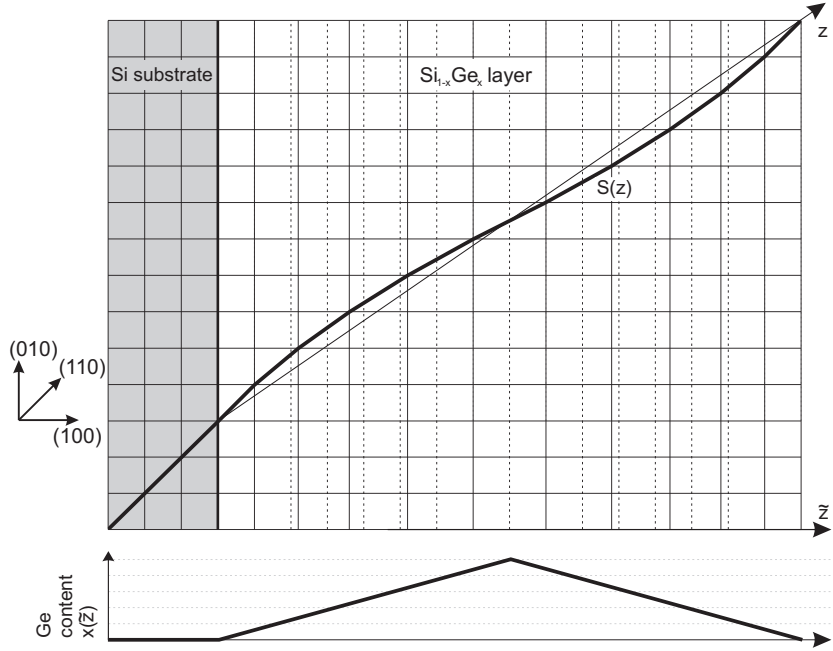}
\caption{Schematic figure of the first period of the bent crystal.}
\label{fig:bent_crystal_first_period}
\end{figure}

Using well-known methods of crystal growing (like molecular beam
epitaxy or chemical vapor deposition, see the references in
\cite{Breese97c}) it is possible to add single crystal layers onto a
substrate. Let us consider a pure silicon substrate on which a
$\mathrm{Si}_{1-x}\mathrm{Ge}_x$ layer is added ($x$ denotes the
germanium content in this layer). The doping with germanium leads to
the enlargement of the lattice constant of the added layer. The strain
due to the lattice mismatch of the substrate and the
$\mathrm{Si}_{1-x}\mathrm{Ge}_x$ layer leads to an increase of the
lattice spacing perpendicular to the surface of the substrate (the
$\tilde z$-direction in figure
\ref{fig:bent_crystal_first_period}). The lattice constant parallel to
the surface remains unchanged.

Prior to discussing the growing of periodically bent channels, let us
summarize the main ideas presented in \cite{Breese97c} that we need
for our description. The spacing between the (100) layers is
$d_{\mathrm{Si}}=1.358$ \AA\ in Si and $d_{\mathrm{Ge}}=1.414$ \AA\ in
Ge. The distance between two $\mathrm{Si}_{1-x}\mathrm{Ge}_x$ layers
is given by $d(x)=d_{\mathrm{Si}}+ \Delta d \cdot x$, where $\Delta
d=d_{\mathrm{Ge}} - d_{\mathrm{Si}}$. In \cite{Breese97c} the critical
thickness of the strained layer is discussed. If the thickness of the
strained layer is larger than the critical value $h_c$, then lattice
defects appear and destruct the channels.

To obtain periodically bent channels, one starts with a pure silicon
substrate and adds $\mathrm{Si}_{1-x}\mathrm{Ge}_x$ layers with
continuously increasing Ge content. This results in bending of the
(110) channels in the direction of the (100) channels. The periodicity
of the shape requires the change of the direction of the bending
toward the (010) channels. This, in turn, can be achieved by reducing
$x$ until it reaches 0. Figure \ref{fig:bent_crystal_first_period}
schematically illustrates the first period of the bent (110) channel.

The last (within the first period) crystal layer consists of pure
silicon, so that the second period can be built up on top of the first
in the same manner. To be captured by the bent channel, the positron
beam should be directed towards the (110) channel of the substrate.

The crystal strain is strongest after half a period, when the
germanium content reaches its maximum. The thickness of the layers
corresponding to half a period needs to be smaller than the critical
thickness $h_c$. If this condition is met, then crystals with
arbitrary number of undulator periods can be constructed.

We now present the formulas that allow to calculate the germanium
content $x(\tilde z)$ as a function of the thickness $\tilde z$ of the
crystal for a given shape $S(z)$.

The differential equation which relates the (local) curvature of the
bent channel and the function $x(\tilde z)$ reads:
\begin{equation}
\label{bent_crystal_11}
\frac{S^{\prime\prime}\flr(\zeta)}
{\lr({S^{\prime 2}\flr(\zeta)+1})^{3/2}}
=
- \frac{d_{\mathrm{Si}}\ \Delta d\ \lr({d_{\mathrm{Si}}+ \Delta d \cdot
x(\tilde z)})} {\lr({d_{\mathrm{Si}}^2 + \lr({d_{\mathrm{Si}}+ \Delta d
\cdot x(\tilde z)})^2})^{3/2}}\ x^\prime(\tilde z),
\end{equation}
where $\tilde z$ is the coordinate in the direction of the crystal
growth, $\zeta=\tilde z/\cos \varphi + S\flr({\tilde z/\cos
\varphi})$ and $\varphi=\pi/4 -
\arctan S^{\prime}|_{z=0}$. The prime denotes the derivative with
respect to the argument.

\begin{figure}
\includegraphics{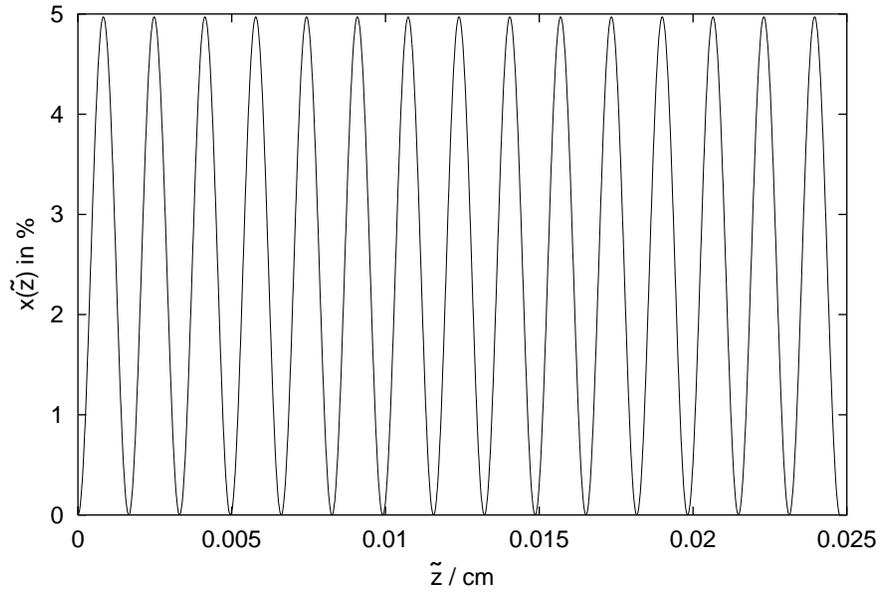}
\caption{Germanium content as function of $\tilde z$ for a bent
Si(110) crystal. The bent channels have the shape $a\, \sin kz$ with
$a=10\,d=1,92\cdot 10^{-7}$ cm and $\lambda=2\pi/k = 2.335\cdot
10^{-3}$ cm. The same parameters are used for the calculation of the
spectrum for $a/d=10$ shown in figure \ref{fig:spectra_500mev}.}
\label{fig:crystal_500mev_x}
\end{figure}

To illustrate the application of equation (\ref{bent_crystal_11}) we
consider two examples. First we discuss growing the Si crystal with
the sine-like shape $S(z)= a\, \sin kz$ with $a=10\,d=1,92\cdot
10^{-7}$ cm and $\lambda=2\pi/k = 2.335\cdot 10^{-3}$ cm.  These
parameters correspond to the undulator emission spectrum presented in
figure \ref{fig:spectra_500mev} for $a/d=10$. The germanium content
obtained by solving numerically the differential equation
(\ref{bent_crystal_11}) is plotted in figure \ref{fig:crystal_500mev_x}.

The maximum germanium content is 5\%. The layer thickness that
corresponds to half a period is given by $\lambda/(2\,\sqrt{2})
=0.8\cdot 10^{-3}$ cm. The critical thickness $h_c$ for a strained
crystal with 5\% of Ge is about $1.2\cdot 10^{-3}$ cm \cite{Breese97c}.

The second example concerns the shape function given by
(\ref{losses_comp_3}) with $a(z)$ and $\lambda(z)$ as in figure
\ref{fig:losses_lambda}. To find the dependence $x(\tilde
z)$ in this case is not so straightforward as for the sine
profile. Indeed, if one starts integrating (\ref{bent_crystal_11})
from $z=0$, then the solution results in negative values of $x(\tilde
z)$. To understand this non-physical result we take a closer look at
(\ref{bent_crystal_11}). For small $x$ this equation acquires the
following approximate form:
\begin{equation}
\label{bent_crystal_11b}
x^\prime(\tilde z)
\approx
- 
S^{\prime\prime}\flr(\zeta)
\,
\frac{\sqrt{8}\, d_{\mathrm{Si}}}
{\Delta d}
\end{equation}
which leads to
\begin{equation}
\label{bent_crystal_11c}
x(\tilde z)
\approx
\frac{\sqrt{8}\, d_{\mathrm{Si}}}{\Delta d}
\,
\int_0^{\tilde z}
a(\zeta)\, \frac{4\pi^2}{\lambda^2(\zeta)}\, \sin\flr({\varphi(\zeta)})
\,
\d \tilde z.
\end{equation}
Using the values $a$- and $\lambda$-values shown in figure
\ref{fig:losses_lambda} one finds that the right hand side can be
negative for some values $\tilde z$.

To avoid this problem one can consider the crystal growth in the
inverse direction: $S(z) \rightarrow S(L_d -z)$ for $0 \le z \le L_d$.
Then the factor $a(\zeta)/\lambda^2(\zeta)$ decreases with
$\zeta$ increasing and the integral in (\ref{bent_crystal_11c}) is
positive for all $\tilde z>0$.

\begin{figure}
\includegraphics{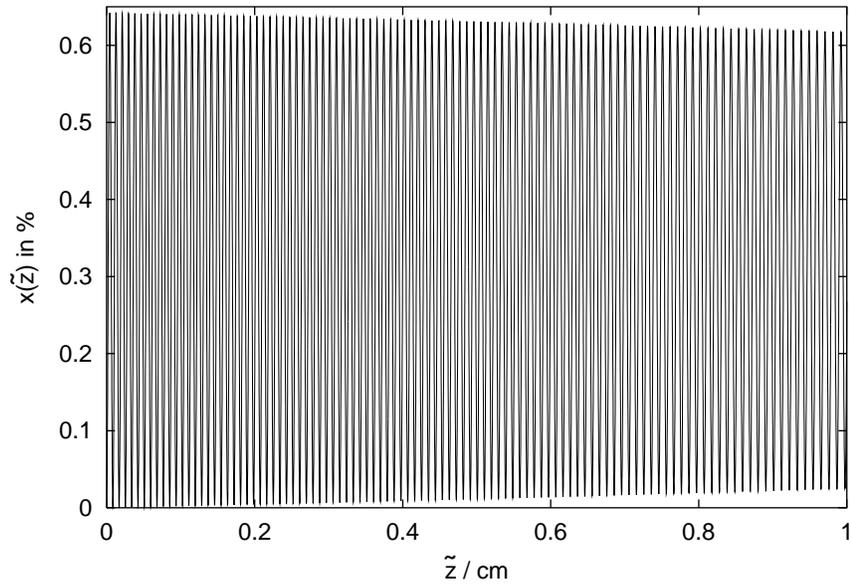}
\caption{Germanium content as function of $\tilde z$ for a bent
Si(110) crystal. The shape of the crystal corresponds to the example
discussed in the previous section.}
\label{fig:crystal_50gev_x}
\end{figure}

The projectiles are injected not through the substrate, as in the
first example, but from the opposite side of the crystal. The results
of the calculation of the germanium content are shown in figure
\ref{fig:crystal_50gev_x}.

The maximal germanium content is smaller than 0.65\% which gives a
critical thickness of $h_c=0.15$ cm (see \cite{Breese97c}). Thus the
critical thickness is much larger than the thickness of the layers:
$\lambda_0/(2\,\sqrt{2}) = 0.8 \cdot 10^{-2}$ cm.  Over the total
length of the crystal (about 3 cm) the minimal Ge content grows
continuously up to $\sim 0.1\%$. The critical thickness for this
Ge content is about 9 cm \cite{Breese97c} which is three times larger
than the length of the crystal.

\section{Summary and outlook}

In this work we have discussed the high-energy regime of the undulator
radiation emitted by ultra-relativistic positrons channeling in
periodically bent crystal channels.

This regime is typical for positron energies well above 10 GeV, when
the channeling effect is accompanied by noticeable radiative
losses. The latter, being mainly due to the channeling radiation, lead
to the gradual decrease of the positron energy. This, in turn,
strongly influences the stability of the parameters of the emission of
undulator radiation.

We demonstrated that the frequency of the undulator radiation can be
maintained constant provided the parameters of the periodic bending
are changed with the penetration distance to take into account the
decrease of the projectile energy.

Our investigation shows that the discussed modification of the shape
of the crystal channels allows the generation of undulator radiation
of high-energy photons (up to tens of MeV). The calculation of the
spectral distributions of the emitted photons in this regime is
currently in progress and will be reported soon.

We described a method that should allow the growing of the
crystal channels that are necessary for the experimental measurement
of the photon spectra. The feedback from experimentalists would
be very helpful to check the models and assumptions that were used in
this work.

\begin{theacknowledgments}
The research was supported by DFG, BMBF and the Alexander von Humboldt
Foundation.
\end{theacknowledgments}

\bibliographystyle{aipproc}
\bibliography{paper_200106}

\end{document}